\documentclass[twocolumn]{revtex4}
\usepackage{graphics,graphicx}
\usepackage{ulem}
\begin{document}
\title{Spin correlation in trigonal EuMn$_{2}$As$_{2}$}
\author{A. Dahal$^{1}$}
\author{Yiyao Chen$^{1}$}
\author{T. Heitmann$^{2}$}
\author{A. Thamizhavel$^{3}$}
\author{S. K. Dhar$^{3}$}
\author{D. K.~Singh$^{1,*}$}
\affiliation{$^{1}$Department of Physics and Astronomy, University of Missouri, Columbia, MO 65211}
\affiliation{$^{2}$Missouri University Research Reactor, University of Missouri, Columbia, MO 65211}
\affiliation{$^{3}$Department of Materials Sciences, Tata Institute of Fundamental Research, Mumbai, India}
\affiliation{$^{*}$email: singhdk@missouri.edu}

\begin{abstract}
The trigonal structure of EuMn$_{2}$As$_{2}$ is an anomaly in the tetragonal 122-type pnictide family. We report detailed investigation of the underlying magnetic correlations in single crystal EuMn$_{2}$As$_{2}$ using high resolution elastic neutron scattering measurements. The system undergoes through two successive antiferromagnetic transitions at $T$ = 135 K and 14.4 K, respectively. Numerical modeling of the experimental data reveals the long range antiferromagnetic correlation of Mn-ions in the $a-b$ plane below $T_N$$_{1}$ = 135 K. Mn spins are aligned closer to the diagonal axis of the unit cell. The lower temperature transition, below $T_N$$_{2}$ = 14.4 K, is found to arise due to the long range antiferromagnetic correlation of Eu spins that are rotated by $\theta$ =55 degree from the $c$-axis of the unit cell. 

\end{abstract}

\maketitle

Electronic interactions in strongly correlated systems affect the physical and magnetic properties of quantum particles that ultimately results in different phases of matter.\cite{Coleman,Sachdev,Varma} Two such novel phases that often compete are superconductivity and magnetism. However, in a set of materials, for example cuprates,\cite{Tranquada} and heavy fermions,\cite{Steglich} magnetism and superconductivity are not only found to coexist but
a parallel resemblance in the energetic interplay between the two are also proposed.\cite{Fisk} Pnictide magnets with 122-type stoichiometric configuration have emerged as a novel platform to explore magnetism and its interplay with superconductivity. Many pnictide materials, such as BaFe$_{2}$As$_{2}$, EuFe$_{2}$As$_{2}$, are known to exhibit unconventional superconductivity where the underlying magnetism is argued to play important role in the Cooper pair formation.\cite{Johnston,Stewart,Scalapino,Dagotto,Fernandes,Hosono} In addition to the interplay between magnetism and superconductivity, pnictides are also at the forefront of the exploration of novel magnetism, such as valence fluctuation and itinerant magnetism.\cite{Schlottmann,Singh1,Anand1,Paramanik,Singh3} Often enough, these materials exhibit multiple magnetic ordered regimes that are intricately related to the underlying lattice structure.\cite{Scalapino2} 122-pnictides usually crystalize in ThCr$_{2}$Si$_{2}$-type tetragonal structure.\cite{Stewart} EuMn$_{2}$As$_{2}$ is an anomaly to this hierarchical crystal symmetry. It crystallizes in trigonal CaAl$ _{2} $Si$ _{2} $-type structure in the space group $P-3m1$ with lattice constants of a = b = 4.23 $\AA$ and c = 7.263 $\AA$.\cite{Anand1, Anand2, Rhul} Careful analysis of the lattice structure reveals an interesting combination of magnetic ion arrangements in EuMn$_{2}$As$_{2}$, see Fig. 1, where Eu ions, S = 7/2, occupy the trigonal sites that are at the center of honeycomb sub-cell of Mn-ions with S = 5/2.\cite{Anand3} We also note that EuMn$_{2}$As$_{2}$ exhibits insulating electrical characteristics, which is unusual of 122-pnictides that are known to exhibit metallic properties.\cite{Stewart,Steglich} The lattice structure and the ensuing lattice-spin interaction seems to be at the core of this puzzling behavior. However, information about the underlying magnetic correlations in EuMn$_{2}$As$_{2}$ is alluding.

\begin{figure}
\centering
\includegraphics[width = 7.6 cm] {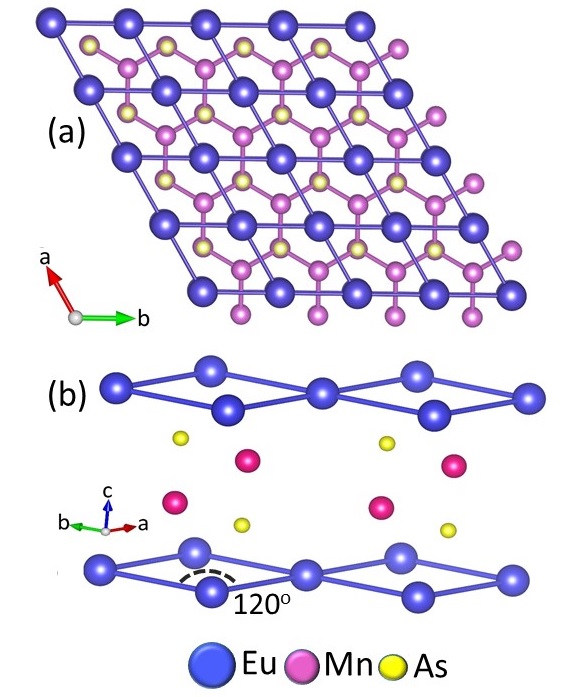} \vspace{-4mm}
\caption{(color online) Crystal structure of EuMn$ _{2} $As$ _{2}$. (a) Mn ions arrange themselves in a two-dimensional honeycomb pattern (described in ref. [21]). Blue, purple and yellow spheres represent Eu, Mn and As ions. (b) CaAl$ _{2} $As$ _{2} $-type trigonal crystal structure of EuMn$ _{2} $As$ _{2} $(space group $P-3m1$). Eu ions arrange themselves on the vertices of trigonal lattice. }
\vspace{-6mm}
\end{figure}

\begin{figure*}
\centering
\includegraphics[width = 17 cm] {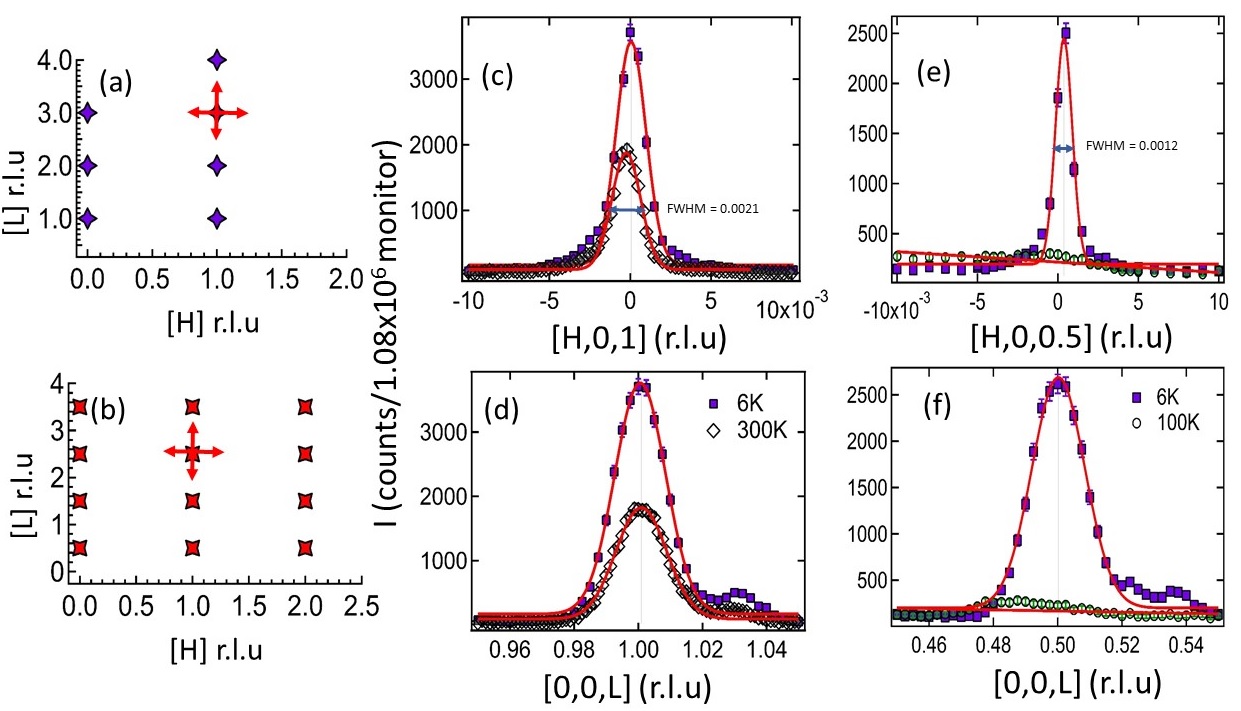} \vspace{-4mm}
\caption{(color online) Characteristic plots of elastic measurement. (a-b) Schematic description of elastic scans along $H$ and $L$ directions across several Brillouin zones.(c-d) Scans along $H$ and $L$ directions across [001] nuclear peak at $\sim$6K and 300K, respectively. Strong enhancement in scattering intensity is observed as sample is cooled to low temperature of $T$ = 6 K. Experimental data are well described by Gaussian lineshape. Elastic scans are instrument resolution limited, thus suggesting the long range nature of spin correlation. (e-f) Elastic scans across [001/2] peak, along $H$ and $L$ directions, at $\sim$6K and 100K are shown here. Unlike the nuclear peaks, the peak completely disappears at higher temperature. Once again, the peaks are instrument resolution limited. Error bars represent one standard deviation.}
\vspace{-5mm}
\end{figure*}

In this article, we report on the detailed elastic neutron scattering investigation of single crystal EuMn$_{2}$As$_{2}$. It is argued to exhibit two successive long range antiferromagnetic transitions at $T_N$ $\simeq$ 140 K and 14 K, respectively.\cite{Anand3} Although, previous magnetic and thermodynamic measurements on EuMn$_{2}$As$_{2}$ suggested the development of Eu-ion correlation below $T_N$ $\simeq$ 14 K, the nature of spin correlation at higher temperature is not clear. In fact, no clear signature of magnetic transition at $T_N$ $\simeq$ 140 K was observed in magnetic measurements on EuMn$_{2}$As$_{2}$. Instead, a small cusp in the heat capacity measurement at $T \simeq$ 140 K was identified as the second transition and ascribed to the ordering of Mn-ions of spin S=5/2.\cite{Anand3} It was indirectly inferred by comparing the thermodynamic properties of EuMn$_{2}$As$_{2}$ to an isostructural compound SrMn$_{2}$As$_{2}$ with only one magnetic site.\cite{Anand2} Using elastic neutron scattering measurements, we find direct evidences to two magnetic transitions at $T_N$1 = 135 K, and $T_N$2 = 14.4 K, respectively. Numerical modeling of neutron data elucidates the nature of spin correlation in both magnetic regimes of T $\leq$ 135 K and T $\leq$ 14.4 K. While the high temperature magnetic phase is found to be associated to the antiferromagnetic correlation of Mn-ions in the $a-b$ plane, the lower temperature magnetic phase is dominated by the long range antiferromagnetic correlation of Eu-spins, with S = 7/2, along the $c$-axis.

Neutron scattering measurements were performed on 10 mg flux grown single crystal of EuMn$ _{2} $As$ _{2} $ at the thermal Triple Axis Spectrometer, TRIAX, at the University of Missouri Research Reactor (MURR). Elastic measurements were performed at the fixed final energy of 14.7 meV. The measurements on TRIAX employed a flat pyrolytic graphite (PG) analyzer with collimator sequence of PG filter-60'-60'-Sample-40'-PG filter-40'. Single crystal sample was mounted at the end of the cold finger of a closed cycle refrigerator with a base temperature of $T \simeq$ 5 K. Measurements were performed with the crystal oriented in the ($H0L$) scattering plane. Here $H$ and $L$ represent reciprocal lattice units of 2${\pi}/a$ and 2${\pi}/c$, respectively.

Single-crystal measurements allow for a detailed examination of the intensities and magnetic scattering pattern, which reveals the nature of spin correlations that are not possible to obtain from magnetic and thermodynamic measurements. Elastic scans were obtained along both $H$- and $L$-crystallographic directions, as shown schematically in Fig. 2a-b. In Fig. 2c-d, we plot background corrected elastic representative scans at $T$ = 6 K and 300 K across the [001] nuclear Bragg peak in $H$ and $L$ directions. As the sample is cooled to low temperature, resolution-limited enhancement of nuclear Bragg peak becomes apparent. Neutron scattering data is well described by the Gaussian line shape. The additional scattering, which is arising due to magnetic correlation in the sample, indicates the development of commensurate long range magnetic order in the system. Interestingly, elastic measurement across [001/2] reciprocal lattice vector also shows the development of resolution-limited magnetic peak as temperature is reduced to $T$ = 6 K, see Fig. 2e-f. It suggests that the ground state magnetic configuration is antiferromagnetic in nature. Low temperature elastic measurements depict magnetic Bragg peaks at the positions, $h+k+l =(n+1)/2$ (shown in Fig. 2a-b).

\begin{figure}
\centering
\includegraphics[width = 8.6 cm] {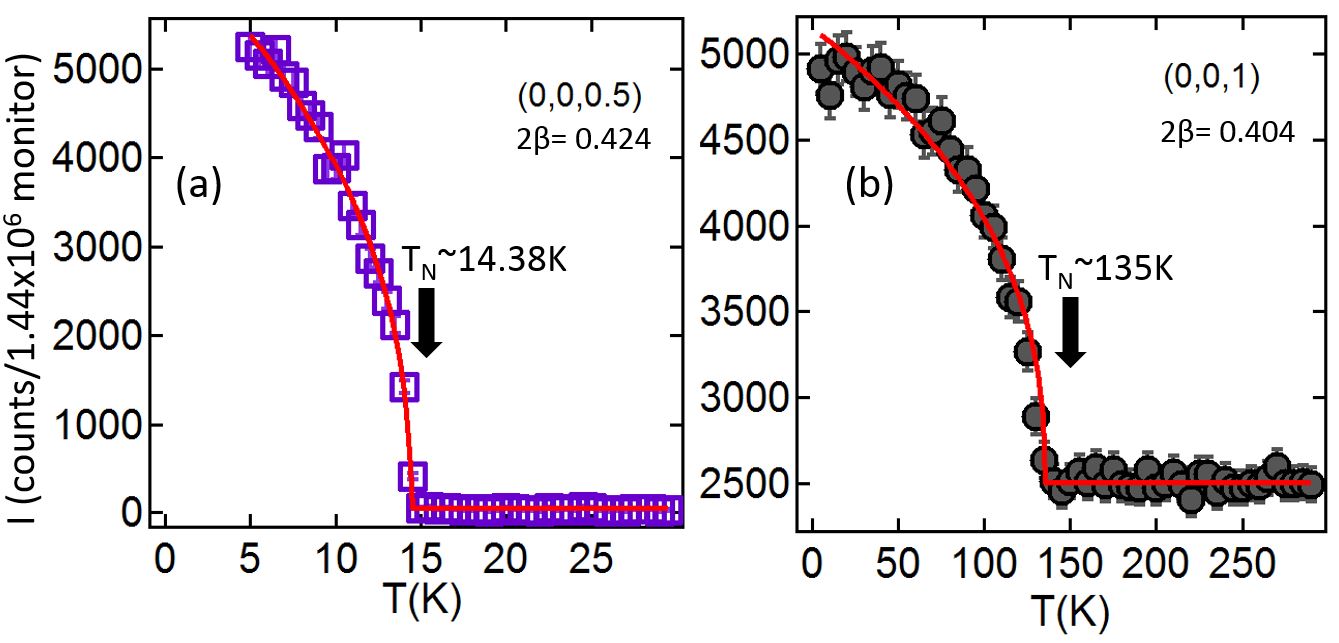} \vspace{-4mm}
\caption{(Color online) Order parameter measurement. (a) Order Parameter measurement performed at [001/2] peak position. Experimental data is fitted with a power law (see text for detail) to obtain information about magnetic transition temperature. The system tends to develop magnetic order below $T$ = 14.38K, with critical exponent 2$ \beta $ = 0.42 (b) Measurement of order parameter at nuclear peak position [001] reveals an entirely different magnetic ordered regime, as manifested by the transition temperature of $T$ = 135K. However, the critical exponent 2$\beta$ is comparable, $\sim$ 0.4, to that observed at [001/2]. Error bars represent one standard deviation.} 
\vspace{-5mm}
\end{figure}

The observation of long range antiferromagnetic order at low temperature in EuMn$ _{2} $As$ _{2} $ is consistent with a previous report, based on magnetic and thermodynamic measurements.\cite{Anand3} However, direct evidence to magnetic transition at $T \simeq$ 140 K is lacking. Elastic scans, shown in Fig. 2c-f, hint of the existence of two different magnetic correlations; one commensurate with the lattice structure and the other with a magnetic unit cell twice that of the lattice unit cell. To know if the two magnetic correlations manifest different temperature dependences, we have performed the measurement of order parameters as a function of temperature at magnetic Bragg peaks of [001] and [001/2]. As shown in Fig. 3, two entirely different temperature dependences are manifested by the system. The order parameter data are fitted with the power law equation: $ I \propto  (1-\frac{T}{T_{N}})^{2\beta} $, to accurately estimate the magnetic transition temperature and the critical exponents that can provide important information about the nature of the phase transition. While low temperature transition is found to occur at $T$ = 14.4K and manifest a critical exponent of 2$\beta$ = 0.42, the higher temperature transition occurs at $T$ = 135K with a critical exponent of 2$\beta$ = 0.40. The value of exponent $\beta$ also suggests three-dimensional interaction in the ordering regime.\cite{vicari} Corresponding ordered moments for integer and half-integer diffraction peaks, at $T$ = 6 K, are determined to be 2.8(3) $\mu_B$ and 1.6(4) $\mu_B$, respectively, which indicates the dominance of Mn-spin at higher temperature and Eu-spin at lower temperature.

\begin{figure*}
\centering
\includegraphics[width = 18. cm] {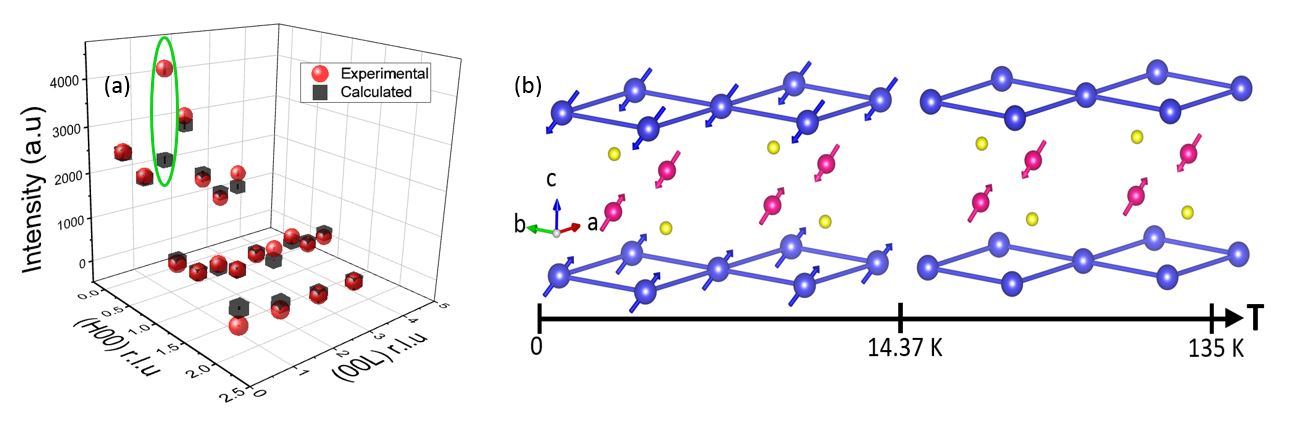} \vspace{-4mm}
\caption{(Color online) Eu and Mn spin correlations in EuMn$ _{2} $As$ _{2}$. (a) Comparison of experimental, obtained at $T$ = 6 K), and calculated intensities at magnetic Bragg peaks. Good agreement between experimental and calculated intensities is obtained for the proposed spin structure. Green ellipse, enclosing a data point and calculated intensity, highlights spurious scattering. Error bars represent one standard deviation. (b) Numerical modeling of neutron data reveals two different antiferromagnetic orders. Mn spins tend to develop antiferromagnetic correlation below $T$ = 135 K in the $a-b$ plane where they are rotated by $\varphi$ = 12 degree with respect to $a$-axis and $\theta$ = 55 degree with respect to $c$-axis. As temperature is reduced below $T$ = 14.4 K, new magnetic order due to Eu spin antiferromagnetic correlation arises. Unlike Mn spins, Eu spins order along the $c$-axis where they are rotated by $\varphi$ = 5 degree in the $a-b$ plane and by $\theta$ = 55 degree with respect to $c$-axis.}
\vspace{-5mm}
\end{figure*}

Clearly, EuMn$ _{2} $As$ _{2} $ exhibit two different magnetic ordered regimes. The transition at higher temperature seems to be arising due to the long range correlation of Mn-spins, which persists to the lowest measurement temperature. A new magnetic order, however, develops below $T$ = 14.4 K, which is most likely associated to Eu-spins. Magnetic orders in the two sub-lattices of Mn and Eu ions are, apparently, cooperative in nature. Insights about the nature of spin correlations in Mn and Eu sub-lattices are gained from numerical modeling of the elastic scattering data. For this purpose, elastic measurement was performed across several Brillouin zones, see FIg. 4a. The experimentally observed structure factor, estimated from the Gaussian fit of the elastic data, are compared with the numerically calculated structure factor for model spin configurations. Structure factor is calculated using, $ F_{M} =\sum_{j} S_{\perp j} p_{j}  e^{ iQr_{j} }e^{-W_{j}} $,\cite{Shirane} where $ S_{\perp} =\hat{Q}\times(S\times\hat{Q})$ is the spin component perpendicular to the Q, $ p = (\frac{\gamma r_{0}}{2})gf(Q) $, $ (\frac{\gamma r_{0}}{2}) $= 0.2695 $ \times 10^{-12} cm$, g is the Lande splitting factor and was taken to be g = 2, $ f(Q) $ is the magnetic form factor, and $e^{-W_{j}} $ is the Debye-Waller factor and was taken to be 1. Since Eu is a strong absorber of neutron, absorption corrections were made to the elastic neutron data. The absorption correction was performed by multiplying the experimental data with the inverse of transmission coefficient $A$. EuMn$ _{2} $As$ _{2}$ single crystal, used in this study, was a very thin sample (thickness $\simeq$ 1.25 mm) with slab geometry. For such a geometry, coefficient $A$ in the case of maximum absorption or negligible transmission is given by,
\begin{eqnarray}
A = \frac{sin(\theta - \varphi)}{\mu \lbrace sin  (\theta - \varphi) + sin (\theta + \varphi)\rbrace}
\end{eqnarray},
where $\mu$ is neutron absorption coefficient, $\theta$ is the sample angle and $\varphi$ is the angle between sample surface and the scattering plane. For Eu, $\mu$ $\simeq$ 140 cm$^{-1}$ at 14.7 meV.\cite{NIST} In a particular situation when neutron beam is parallel to the scattering plane of the crystal such that the absorption is higher i.e. $\varphi$ = 0, the coefficient A simplifies to, A = [1 - exp(2$\mu$t.cosec$\theta$)]/2$\mu$. Absorption correction to the diffraction data did not make any significant difference for such a thin sample. Correction to the estimated intensities due to Eu absorption were found to be within the statistical error margin. Numerical modeling was performed for various different scenario. Best fit to the neutron diffraction data at integer positions is obtained for a spin configuration where Mn-moments are arranged antiferromagnetically in the $a-b$ plane and aligned ferromagnetically along the $c$-axis. In this process, moments are rotated by $\theta$ = 55$^{o}$ with respect to the $c$-axis and $\varphi$ = 12$^{o}$ with respect to $a$-axis in the $a-b$ plane, see Fig. 4b. Lower transition temperature magnetic phase, represented by diffraction peaks at half-integer positions, is well described by antiferromagnetic spin correlation of Eu spins along the $c$-axis where moments are rotated by $\theta$ = 55$^{o}$, with respect to $c$-axis, and $\varphi$ = 5$^{o}$ with respect to $a$-axis in the $a-b$ plane. 

In summary, we have investigated the nature of spin correlation and magnetic transition in single crystal EuMn$ _{2} $As$ _{2}$. The transition at low temperature, $T_N$ = 14.4 K, is somewhat consistent with previous magnetic and thermodynamic measurements. But the second transition at higher temperature, $T_N$ = 135 K, was not observed before. Numerical modeling of the elastic neutron data reveals distinct long range antiferromagnetic correlation of Eu and Mn spins. While Mn spins are antiferromagnetically aligned in the $a-b$ plane, Eu spins develop AFM order along the c-axis. However, in both cases, moments are significantly rotated from the $c$-axis and aligned closely to the diagonal axis. The modeled spin configuration is different from the previously proposed Ising configuration in this system.\cite{Anand3} Many pnictide materials of lanthanide family are known to exhibit multiple magnetic transitions.\cite{Wolfe,Singh2} Neutron based study of EuMn$ _{2} $As$ _{2}$ will help us in developing a comprehensive understanding of this complex behavior. Future research works, especially inelastic neutron scattering measurements on large single crystal of EuMn$ _{2} $As$ _{2} $, are highly desirable. Understanding the interaction between Eu and Mn spins as well as the valence fluctuation of Eu ions can shed light on the cooperative nature of spin configurations and their role in the anomalous insulating electrical properties in this system. 

The research at MU is supported by the U.S. Department of Energy, Office of Basic Energy Sciences under Grant number DE-SC0014461. Authors also thank J. Gunasekera for help with neutron scattering experiments.

\clearpage


\begin{thebibliography}{99}


\bibitem{Coleman} P. Coleman and A. Schofield, \textit{Nature} \textbf{433}, 226-229 (2005).

\bibitem{Sachdev} S. Sachdev, Quantum Criticality, Cambridge Univ. Press, New York, 1999.

\bibitem{Varma} C. M. Varma, \textit{Rev. Mod. Phys.} \textbf{48}, 219-238 (1976).

\bibitem{Tranquada} J. M. Tranquada,  H. Woo, T. G. Perring, H. Goka, G. D. Gu, G. Xu, M. Fujita and K. Yamada, \textit{Nature} \textbf{429}, 531 (2004).


\bibitem{Steglich} P. Gegenwart, Q. Si and F. Steglich, \textit{Nature Physics} \textbf{4}, 186 (2008).

\bibitem{Fisk} Z. Fisk, D. W. Hess, C. J. Pethick, D. Pines, J. L. Smith, J. D. Thompson and J. O. Willis, \textit{Science} \textbf{239}, 33 (1988)

\bibitem{Johnston} D.C. Johnston, \textit{Adv. Phys.} \textbf{59}, 803 (2010).
\bibitem{Stewart} G.R. Stewart, \textit{Rev. Mod. Phys.} \textbf{83}, 1589 (2011).
\bibitem{Scalapino} D.J. Scalapino, \textit{Rev. Mod. Phys} \textbf{84}, 1383 (2012).
\bibitem{Dagotto} E. Dagotto, \textit{Rev. Mod. Phys.} \textbf{85}, 849 (2013).
\bibitem{Fernandes} R.M. Fernandes, A. V. Chubukov, and J. Schmalian, \textit{Nat. Phys.} \textbf{10}, 97 (2014).

\bibitem{Hosono} H. Hosono and K. Kuroki, \textit{Physica C} \textbf{514}, 399 (2015).

\bibitem{Anand1}V. K. Anand, D. T. Adroja, A. Bhattacharyya, U. B. Paramanik, P. Manuel, A. D. Hillier, D. Khalyavin, and Z. Hossain, \textit{Phys. rev. B} \textbf{91}, 094427 (2015).

\bibitem{Schlottmann} P. Schlottmann, \textit{Phys. Rev. B} \textbf{75}, 205108 (2007).

\bibitem{Singh1} D. K. Singh, A. Thamizhavel, S. Chang, J. W. Lynn, D. A. Joshi, S. K. Dhar, and S. Chi, \textit{Phys. Rev. B} \textbf{84}, 052401 (2011).

\bibitem{Paramanik} U. B. Paramanik, R. Prasad, C. Geibel, and Z. Hossain, \textit{Phys. Rev. B} \textbf{89}, 144423 (2014).

\bibitem{Singh3} D. K. Singh and M. T. Tuominen, \textit{Phys. Rev. B} \textbf{83}, 014408 (2011).

\bibitem{Scalapino2} S Graser, T A Maier, P J Hirschfeld and D J Scalapino, \textit{New J. Phys.} \textbf{11}, 142 (2009).


\bibitem{Anand2}V. K. Anand, P. Kanchana Perera, Abhishek Pandey, R. J. Goetsch, A. Kreyssig, and D. C. Johnston, \textit{Phys. rev. B} \textbf{85}, 214523 (2012).
\bibitem{Rhul}R. Ruhl and W. Jeitschko, \textit{Mater. Res. Bull.} \textbf{14}, 513 (1979).
\bibitem{Anand3}V. K. Anand and D.C. Johnston, \textit{Phys. rev. B} \textbf{94}, 014431 (2016).



\bibitem{vicari} A. Pelissetto and E. Vicari, \textit{Phys. Rep.} \textbf{368}, 549 (2002).

\bibitem{Shirane} G. Shirane, J. Tranquada and S. Shapiro, \textit{Neutron Scattering with a Triple Axis Spectrometer: Basic Techniques}, Cambridge University Press, New York, 2002.

\bibitem{NIST} Information about neutron absorption coefficient can be found at https://www.ncnr.nist.gov/instruments/bt1/neutron.html.

\bibitem{Wolfe} H. Lohneysen, A. Rosch, , M. Vojta, and P. Wolfe, \textit{Rev. Mod. Phys.} \textbf{79}, 1015 (2007).

\bibitem{Singh2} D. K. Singh, A. Thamizhavel, J. W. Lynn, S. K. Dhar, and T. Hermann, \textit{Phys. Rev. B} \textbf{86}, 060405(R) (2012).
\\



\end{thebibliography}
\end{document}